# Techniques for Securing Data Exchange between a Database Server and a Client Program

**Ovidiu Crista**
**"Tibiscus" University of Timişoara, România**

ABSTRACT. The goal of the presented work is to illustrate a method by which the data exchange between a standalone computer software and a shared database server can be protected of unauthorized interception of the traffic in Internet network, a transport network for data managed by those two systems, interception by which an attacker could gain illegetimate access to the database, threatening this way the data integrity and compromising the database.

## 1. Generalities

In client-server applications securing data exchange between two systems is a very difficult task, considering a possible intrusion of an attacker on the flow of data. The real risk is that one could gain access to the server by listening to the network traffic and clone the data transmitted with modified parameters and have the server behave according to attacker's wish.

The model presented here can be implemented as a communication protocol between a Windows based computer application written in C++ and any database server that provides an application programming interface for the PHP scripting language.

As an alternative to other specification protocol for exchanging structured information like SOAP for example, which relies on XML, this model has the advantage to reduce the amount of format messages as values of data are delimited only by an ASCII character. If in an XML-based protocol is used, the transmitter or receiver of data should also implement routines for parsing the XML code.

95

## 2. The working of the protocol

The protocol has been designed to carry data over the Internet between two computers and is build on the top of the HTTP application-level protocol.

The client program using the protocol would initiate and prepare http requests for sending them to a predefined server which in its turn, parses and processes the data transmitted and finally sends back the results.

Most times the https requests lead to an SQL query executed on the database server and the client program does not need to know how a query is made. Every SQL command is stored as a string constant with placeholders for variables received from client programs.

For client programs, there is a C++ class developed to achieve the http requests' behaviors for sending data for processing. On the other hand, the server uses a web server program to receive the https requests. With help of the PHP scripting language the requests are read which produces a data structure for the scripts running on server.

## 3. Data transmission

Whenever the client program needs some data from server or it has to send some values to be stored in the database, it has to prepare an http request with the help of the C++ class defined as "CApelServer", a class compiled together with all other classes that make the resulting executable.

This wrapper class makes use of other already available classes from the Microsoft Foundation Classes framework:
- CInternetSession to create and initialize an Internet session for parsing URL's and managing http connections;
- CHttpConnection to create the connection to the server;
- CHttpFile for sending the request and receive the data as an ordinary file.

The "CApelServer" class takes all the responsibility to manage connections, to prepare data and to send the HTTP requests.

When the client program has enough information for making a request, probably as a result of a multiple user commands, an object of type CApelServer is constructed. Then, through a function call the http request is made.

Example:

```
#define HOST _T("tibgs.tibiscus.ro")
#define URL _T("/note/index.php")
#define IDC_COMMAND1 105

CApelServer as();
as.Call(IDC_COMMAND1, (int)StudentID, (CString)sNota);
```

The first two constants defined with preprocessor directives have to be available to the object "as" since they will be necessary in the call of Call function to make the connection with the server located at the specified domain address.

The last constant is a predefined constant with some value and a correspondent in a string resource file for describing the data types of arguments passed to the function Call(). The value of IDC_COMMAND1 will be sent along with the request to identify the command that has to be executed on the server. For the call of the function Call() to work, the string referred by the first parameter should only contain format flags for the next arguments to be included in the request. In the example described above the resource string should be: "%d%s" for an int variable and a pointer to an array of chars. (The CString class has overloads for pointers to TCHAR data types). The member function Call() is a multiple arguments function.

In the body of the function Call() arguments are read and converted to strings if they are not already strings. Then, a POST stream is created as a buffer of ASCII characters. If strings passed are composed from wide chars or Unicode chars, they are also converted to multi byte strings before included in the POST buffer. All the arguments included in the POST stream are separated by the ASCII unit separator character but the succession of arguments has to be kept to match the succession expected by the receiving server.

Example of a request:

```
POST /note/index.php?p=cmd&ver=1&PHPSESSID=6dlueq5 HTTP/1.1
Content-Type: application/x-www-form-urlencoded
User-Agent: TibGSNote
Host: tibgs.tibiscus.ro
Content-Length: 12
Cache-Control: no-cache

105□78541□9□
```

Prior to sending the request, an encrypted connection with the server has to be made. This is accomplished at "CApelServer" object construction when preparing the "CHttpConnection" object used internally. Even if someone sniffs the network at the time of request, he could not understand the piece of text making the request and has no chance to rebuild the request with his own input parameters. This makes the client program to have full control on what is sent to the server.

## 4. Data receiving by the server

As an ordinary website, the PHP scripts that make the server framework are called and parsed by a PHP engine loaded in a web server instance. Data transmitted as a POST string is available to scripts as a PHP input stream (php://input). Before any process over the received data, the server framework checks the command id to make sure that command is allowed to the currently logged user. Depending on configuration, some commands could need authentication while others could be freely available (checking the framework version for example).

If all criteria are satisfied, the input stream is sent to the processing script which first checks the values for the right type. This is necessary to ensure the integrity of pending SQL queries, in which the received values would be included. If someone, inspecting how the client program runs, clones the requests, the server framework would have a chance to finish execution before some inadequate values get in the database.

The framework loads from a configuration file on the server a user defined set of SQL statements identified by the command ids and then replaces the placeholders found in the selected statement with corresponding values received through the PHP input stream. When done, it executes the query against the database server expecting the result.

The results of queries, if any, should be sent back to the client application through the same protocol, a protocol through which the client can take the values from the result and build a new data structure to use. Any result of a SQL query that returns rows of data should consist of a data table in which field values are separated with the unit separator ASCII character while the rows are separated with the record separator. The framework does not have to concern about the type of query (returning rows or not) since the caller knows better what to expect.

The server framework has no ability to contact the connected program client since this on does not implement network listening.

## 5. Data receiving by the client

The client can receive data from server only after a valid request. The data is received back through an http response respecting the model of the protocol. Now that the execution is back in the client program, the "CApelServer" class has to manage the response in order to make data received usable for the main thread of the application. It takes the response stream and converts it in a table data structure so the program can access the rows of data by a row at a time or by accessing a specific field within a specific row.

The client program has the ability to compute the exact number of rows and fields received. It accesses values using pointers to strings. In the table, all values are represented as strings, and the memory allocated for these will be freed automatically after instance of "CApelServer" goes out of scope. Before letting that happen, the code of the program should store the values in the desired form for later use. For example if a query returns a list of products with their corresponding ids, the client could set the names of the products as items texts in a Combo control of a GUI application and the ids as items data. Then, whenever the user makes a selection, the item data of the selection is the id of the selected product.

## 6. Conclusions and improvements for the future

This model has been integrated in software internally used at "Tibiscus" University of Timisoara. There are two major projects which implement this protocol. We own a MySQL database server for storing financial and academic situations of students. Departments of the University responsible with managing these situations are offered standalone programs developed in C++ using Microsoft Foundation Library to connect to data server and use the database. The communication protocol ensures us that transaction is made safe.

Since users connect remotely to the data server the framework existing on the server also manages the user logging. The client program has the responsibility to ask and send to server the user's credentials for authentication. If the logging operation succeeds a session will be started and will be maintained as long as there are http requests, so each request renews the session and keep it alive.

While this framework provides a way for client applications to connect independently and indirectly to a database server, future extensions based on the same communication model could provide programmers a simpler way to develop software which interact each other or use a shared resource located on the web in a safer manner as data carried through Internet is SSL encrypted and the identity of a user is permanently verified to prevent intruders to alter communications or usurp the rights of a legitimate user.